\documentclass[10pt,conference]{IEEEtran}

\usepackage{fancyhdr}
\usepackage[normalem]{ulem}
\usepackage[sort,nocompress]{cite}
\usepackage{acro}
\usepackage{booktabs}
\usepackage[enable]{easy-todo}
\usepackage{mathtools}
\usepackage{physics}
\usepackage{amsmath,amssymb,amsfonts}
\usepackage{algorithmic}
\usepackage{graphicx}
\usepackage{textcomp}
\usepackage{xcolor}
\usepackage{tikz}

\usepackage{subcaption}

\usepackage[bookmarks=true,breaklinks=true,letterpaper=true,colorlinks,citecolor=blue,linkcolor=blue,urlcolor=blue]{hyperref}

\title{Moveless: \underline{M}inimizing QEC \underline{O}verhead on QCCDs via \underline{V}ersatile \underline{E}xecution and \underline{L}ow \underline{E}xcess \underline{S}huttling}
\author{\IEEEauthorblockN{Sahil Khan}
\IEEEauthorblockA{\textit{Duke University} \\
Durham, USA \\
sahil.khan@duke.edu}
\and
\IEEEauthorblockN{Suhas Vittal}
\IEEEauthorblockA{
\textit{Georgia Tech}\\
Atlanta, USA \\
suhaskvittal@gatech.edu}
\and
\IEEEauthorblockN{Kenneth Brown}
\IEEEauthorblockA{
\textit{Duke University}\\
Durham, USA \\
kenneth.r.brown@duke.edu}
\and
\IEEEauthorblockN{Jonathan Baker}
\IEEEauthorblockA{
\textit{Univerity of Texas at Austin}\\
Austin, USA \\
jonathan.baker@austin.utexas.edu}
}

\begin{document}
\date{}
\maketitle

\thispagestyle{empty}

\begin{abstract}
 One of the most promising paths towards large scale fault tolerant quantum computation is the use of quantum error correcting codes. The majority of popular codes are stabilizer codes, in which error information is extracted via the repeated execution of a set of commuting stabilizer measurement circuits and then decoded. Just like every other quantum circuit, these circuits must be compiled to hardware in a way to minimize the total physical error introduced into the system, for example either due to high latency execution or excessive gates to meet connectivity limitations of the target hardware. However, unlike arbitrary quantum circuits, all syndrome extraction circuits have several common properties, for example they have a bipartite connectivity graph, consist only of commuting subcircuits, and have a set of ancilla which are indistinguishable between measurement rounds, among others. For the most part, compilation methods have aimed at being generic, able to map any input circuit into executables on the hardware, and therefore cannot appropriately exploit these properties and result in executables which have higher physical error. In the case of modular trapped ion systems, specifically QCCDs, this corresponds to the insertion of excessive shuttling operations necessary to realize arbitrary qubit interactions. 

We propose a compilation scheme explicitly tailored for the structural regularity of quantum error correction code physical circuits, with the primary objective of minimizing circuit latency from shuttling operations. Our compiler's success is predicated on several key observations: 1. only ancilla or data (but not both) should be shuttled, 2. stabilizers can be executed in any order meaning we can dynamically modify circuit execution on a per-cycle basis 3. ancilla are indistinguishable meaning any can be selected to begin a stabilizer measurement and since only ancilla are shuttled, we retain a fixed-point mapping between cycles, and 4. QCCD hardware limits the number of parallel operations equal to the number traps in the system bounding the rate of stabilizer measurement meaning fewer ancilla are necessary and can be easily reused. Our resulting compiler, leads to QEC circuits which are on average 3.38$\times$ faster to execute (by up to 5.24$\times$) and therefore incur lower physical error. For codes with good decoders, e.g. the color code and surface code, these improvements lead to up to two orders of magnitude of improvement in logical error rates with realistic physical error rates.

\end{abstract}


% Stats for early machine range (Distance 7, T1=5e7,t2=5e6): (Cyclone error correction, Baseline NOT)
%     x axis [0.0001, 0.00025, 0.0005, 0.00075, 0.001]
%     Baseline [0.00011813, 0.00015851, 0.00026709, 0.00041869, 0.0006222]
%     Cyclone [3.04e-06, 6.66e-06, 1.779e-05, 3.931e-05, 7.61e-05]

%     Stats for long term machine range (Distance 7, T1=1e8, T2=1e7) (Both have error correction)
%     real x [0.0001, 0.00025, 0.0005, 0.00075, 0.001]
%     Cyclone [3.3e-07, 9.4e-07, 4.67e-06, 1.288e-05, 2.987e-05]
%     Baseline [9.18e-06, 1.821e-05, 4.399e-05, 8.331e-05, 0.00014778]

\section{Introduction}

Quantum error correction (QEC) is among the most promising method for enabling the execution of programs with exponential speedups on many promising, yet error-sensitive, applications~\cite{shor1999polynomial, kivlichan2020qsimelectrons, childs2018firstqsim, harrow2009hhl, reiher2017nitrogenfixation}. \textit{Fault-Tolerant Quantum Computers (FTQCs)} implement quantum error correction by leveraging physical qubits to encode \textit{logical qubits} that are resistant to error~\cite{kitaev1997toriccodes, landahl2011colorcodes, shor1995qec}. Provided that the physical error rate ($p$) is sufficiently low (e.g., 0.1\%), the logical error rate can be suppressed with increasing redundancy, or \textit{code distance} ($d$).

\begin{figure*}
        \centering
        \includegraphics[width=\linewidth]{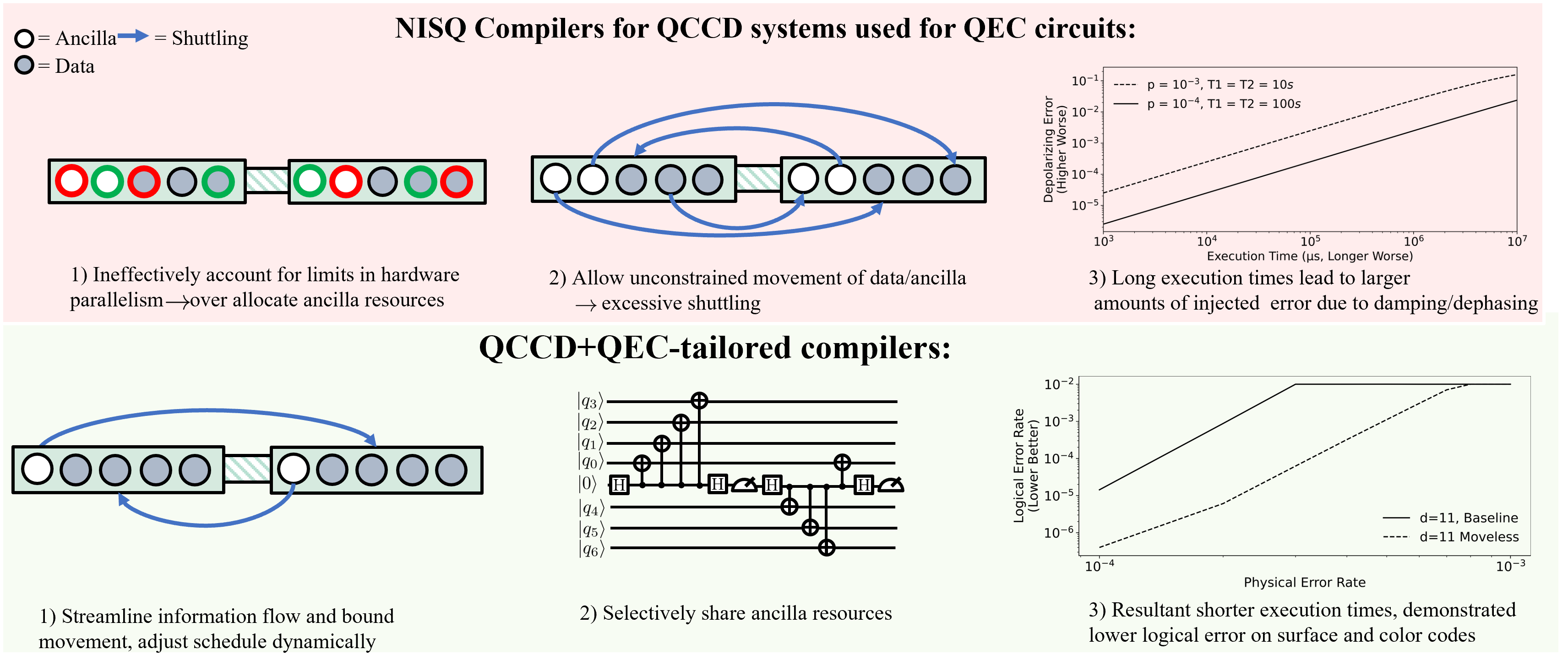}
        \caption{(Top) Historically, most compilers have been generic, i.e. they take in arbitrary quantum circuits and optimize them for a specific target hardware. Because of their generality, they cannot explicitly optimize for the regular properties of QEC circuits. Consequently, these compilers result in only partially optimized circuits with longer than optimal execution times. In the case of QEC, longer circuit duration implies more physical error and therefore worse performance. (Bottom) Our work focuses on optimizing compilation for stabilizer circuits for QCCD systems, exploiting features of both the shared circuit structure (e.g. bipartite connectivity graphs, stabilizer reordering, and circuit reordering, among others) and the target hardware (e.g. limited in-trap parallelism) to substantially reduce shuttling overheads which leads to faster circuits, thus lower physical error, and ultimately lower logical error rates.}
        % A distance three color code in its 3 colorable graph representation is shown on the left. Each face denotes an X and a Z check among its corresponding data qubits indicated by the vertices. On the contrary, a distance 3 surface code with weight 4 and 2 checks, each face represents a different stabilizer check. The circuit form for a surface code check is shown after
        
        \label{fig:fig1}
    \end{figure*}

Quantum error correcting codes encode quantum information across multiple \textit{data qubits}. If left unprotected, these data qubits will incur errors, such as decoherence or heating errors, subsequently spoiling the encoded information. Protecting data qubits requires detecting errors by measuring the parity checks, or \textit{stabilizers}, of the quantum code. To measure the stabilizers of a quantum code, the FTQC must periodically execute a \textit{syndrome extraction circuit}, which entangles the data qubits with one or more \textit{ancilla} qubits, and subsequently measures the ancilla qubits to obtain a bitstring known as a \textit{syndrome}. A classical decoder analyzes the syndrome to identify any errors that have occurred on the logical qubit, which are then corrected. The syndrome extraction circuit itself is faulty, so the syndrome can have false outcomes caused by operation (i.e., CNOT and measurement) errors. In practice, decoders must collectively analyze several consecutive rounds of syndromes to accurately identify errors on data qubits. The rate at which a decoder can accurately correct the errors on these data qubits is known as the \textit{logical error rate}.

In this paper, we consider \textit{Quantum Charge Coupled Devices (QCCD)} trapped ion systems, which are widely considered to the be among the most scalable variant of trapped ion quantum computers \cite{qccdinvention, ryananderson2024highfidelityfaulttolerantteleportationlogical, Lekitsch_2017}. QCCD systems contain several traps of ions, in which qubits have all-to-all connectivity within each trap. However, while each trap supports all-to-all connectivity, they can only perform one or two operations concurrently per trap and have a limited number of total traps \cite{qccdinvention, oxfordionicsparallelism, brown2016codesigningscalablequantumcomputer, honeywellh2}. To perform operations between traps, and maintain system-wide all-to-all connectivity, qubits must be shuttled via inter-trap junctions. The more shuttling that is performed, the more physical error that will be introduced due to decay and so ideally, we shuttle as little as possible \cite{demonstrationQEC}.

\vspace{0.05in}
\noindent
\textbf{Syndrome Extraction in QCCD Systems.} When implementing quantum error correction on quantum computers, there are three factors that determine the feasibility of the code: 
\begin{enumerate}
%[leftmargin=0.5cm,itemindent=0.4cm,labelwidth=\itemindent,labelsep=0.05cm, align=left, itemsep=0.1cm, listparindent=0.5cm, topsep=0.1cm]

\item \textit{Syndrome extraction circuit}: The measurement of the error syndrome circuit will fail depending on its exact implementation which depends on circuit order and timing.

\item \textit{Decoder}:  The information from the syndrome extraction circuit is fed to a decoder. The errors are corrected based on the decoders guess of the error. Decoders have tradeoffs between reliability and speed.

\item \textit{Threshold}: If the syndrome extraction circuit failure rate is low enough for the decoder to fix errors, the logical error rate will improve with increasing code distance for a code family. It is common to parameterize all errors with a single parameter $p$.
    % \item \textit{The threshold of the code}. A code with a poor threshold (i.e. less than 0.1\%) will be very sensitive to noise.
    % \item \textit{The fidelity of the syndrome extraction circuit.} If a syndrome extraction circuit has poor fidelity, then even codes with high threshold will not be useable.
    % \item \textit{The effectiveness of the decoder.} A decoder that cannot properly correct all errors limits the effectiveness of the underlying quantum code.
\end{enumerate}
In QCCD systems, syndrome extraction is nontrivial due to shuttling overheads and limited parallelism. These factors can cause syndrome extraction circuits to be deep, and deep syndrome extraction circuits have poor fidelity as they cause data qubits to be more susceptible to heating, amplitude damping, and dephasing error. Ideally, syndrome extraction circuits should be as short as possible to minimize error. 

% \begin{figure*}[!htb]
%      \centering
%     \includegraphics[width=\linewidth]{asplos25-templates/figures/Figure1Placeholder.pdf}
%      \caption{A syndrome extraction circuit would typically use all ancilla, assigning one ancilla per stabilizer. However, our MMO (minimize movement operations) compiler functions better with lower amounts of ancilla that are reused. This demonstrates a higher threshold, closer to the ideal latency of syndrome extraction, but is not quite ideal. To achieve ideal levels of parallelism, we introduce Cyclone. }
%      \label{fig:intro_fig}
%  \end{figure*}

\noindent
\textbf{Prior Work and Their Limitations.} Existing work on QEC for trapped ions includes small experimental demonstrations \cite{demonstrationQEC, more_experiment_QEC}, as well as schemes on singular traps with large capacity \cite{longchainQEC}. Existing compilation for QCCD systems has been focused on the \textit{Noisy Intermediate Scale Quantum (NISQ)} paradigm \cite{murali1, Saki_2022}, where quantum circuits are run on noisy physical qubits. As these compilers are optimized for NISQ program execution, they are unoptimized for compiling syndrome extraction circuits. 
% For instance, our evaluations of a baseline QCCD compiler \cite{murali1} demonstrates that it results in syndrome extraction circuits which are \textbf{XX} longer  achieves a logical error rate {\color{red} \textbf{XX} higher than the idealized logical error rate (at a set physical error, p}.

Existing compilers are not sufficient for syndrome extraction as they add unnecessary yet costly intertrap movement in three ways. First, existing compilers do not distinguish between data and ancilla qubits in the syndrome extraction circuit, when in principle, a compiler only needs to shuttle either data qubits or ancilla qubits, but not both. Arguably, shuttling \textit{both} data qubits and ancilla qubits, as is done in existing NISQ compilers, is suboptimal because it increases the possibility for unnecessary movement and ``pulling" (where data qubits follow data, and ancilla follows ancilla), increasing overall movement between traps. Second, existing compilers assume static constructions of QEC circuits, e.g. a strict ordering of both stabilizers and their component gates, which is unnecessary. Finally, existing compilers over-allocate ancilla qubits. Because syndrome extraction circuits allow for the reset and subsequent reuse of ancilla qubits, reducing the amount of ancilla in a circuit could limit the overall movement while still maintaining circuit structure. While some prior work has some explicit ancilla sharing (e.g. color codes reusing the ancilla for both X and Z checks), this is generally not the case.

To evaluate the effects of movement minimization and dynamic scheduling on syndrome extraction depth, we combined a shuttling ancilla-only policy, an efficient and adaptive scheduler, along with a full ancilla-reuse policy to create a compiler we designate as \textit{Moveless} explicitly designed to minimize execution time of syndrome extraction circuits. Our evaluations of Moveless show that we reduce circuit depth by up to around 5.24$\times$ and improve logical error rate by around 1-2 orders of magnitude (such as $p = 10^{-4}, d = 11$ surface code as shown in Figure \ref{LER:SCandCC}).  

\vspace{0.05in}
\noindent

% We leverage the key insight that while QCCD systems have low intra-trap parallelism, they can perform multiple concurrent operations, provided each operation occurs in different traps.  

% This dependency of parallelism on underlying hardware hints towards the fact that there could exist a construction of a software/hardware codesign that both achieves bounded overall movement and ideal parallelism.    

% However, amongst common QCCD architectures, which have mostly linear connectivity between traps, operations effectively become serialized {\color{blue} as the compiler is unaware of how to utilize the parallel structure of syndrome extraction circuits on the underlying hardware}, so using more ancilla qubits does not improve parallelism. {\color{blue} Because of this, we postulate that} improving syndrome extraction latency cannot be done solely in software: we need a QCCD architecture that enables parallel execution of operations across multiple ancilla qubits {\color{blue} at the same time as bounding overall intertrap movement. }

% Towards this goal, we make two insights. \textit{First}, while QCCD systems have low intra-trap parallelism, they can perform multiple concurrent operations, provided each operation occurs in different traps. \textit{Second}, {\color{blue} An architecture that supports cyclic shuttling} Given these two insights, we ideally want to spread spatially local data qubits across multiple traps to perform concurrent operations across different traps. 
\begin{figure*}
        \centering
        \includegraphics[width=\linewidth]{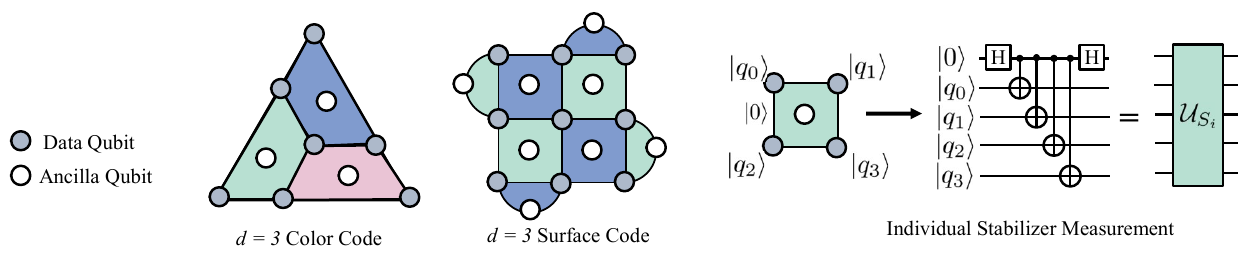}
        \caption{Stabilizer codes are composed of some combination of data qubits and ancilla qubits. Two common code families, the ones considered in this work for logical error rate evaluation, are color codes and surface codes (examples of distance 3 codes are shown here). These codes are composed of a number of stabilizers, given as the colored regions, which define the types of checks used to recover syndrome information. Each check has a corresponding quantum circuit and measurement of the ancilla; an example circuit is given on the right. The weight of the check is determined by the number of data on the boundary of each colored region. After each measurement of the stabilizer the ancilla is reset.  
        % A distance three color code in its 3 colorable graph representation is shown on the left. Each face denotes an X and a Z check among its corresponding data qubits indicated by the vertices. On the contrary, a distance 3 surface code with weight 4 and 2 checks, each face represents a different stabilizer check. The circuit form for a surface code check is shown after
        }
        \label{fig:backgroundcodes}
    \end{figure*}

Our contributions are as follows:
\begin{enumerate}
%[leftmargin=0.5cm,itemindent=0.4cm,labelwidth=\itemindent,labelsep=0.05cm, align=left, itemsep=0.1cm, listparindent=0.5cm, topsep=0.1cm]
    \item We identify that existing NISQ compilers for QCCD systems are suboptimal for QEC circuits as they do not distinguish between data and ancilla qubits.
    \item We detail a technique to dynamically reorder a QEC circuit's stabilizers and relative ordering of gates within each stabilizer according to the changing movement patterns in a circuit.
    \item We find the counterintuitive result that due to limitations in parallelism on near term QCCD systems, ancilla reuse could be more temporally optimal
    \item We model our compiler's effects on logical error rate on surface and color codes, demonstrating orders of magnitude of improvement under realistic noise models.
    
\end{enumerate}

\section{Background and Motivation}

\subsection{Basics of Quantum Error Correction}
\subsubsection{General Stabilizer Codes}
\textit{Stabilizer codes} are part of a broad class of QEC codes, encompassing some of the most popular types of codes in QEC (such as Color and Surface Codes) \cite{gottesman1997stabilizer}. Stabilizer codes can be concisely represented as a set of \textit{stabilizers} ${S_1, S_2, ... S_m}$, where each $S_i$ contains a string of $\{X_j\}$ and/or $\{Z_k\}$ for different data in the data qubits, $n$. The parity between each of these stabilizers are then checked using ancilla qubits. Each check is originally assigned a unique ancilla as to maximize idealized parallelism, but we will revisit this assumption later in Section \ref{subsec:ancilla_reuse}. Stabilizer codes do not have unique representations as they are generated by a set of stabilizer generators. In this work, we will use one example representation of the code, but our designs will work for \textit{any} representation of these codes, such as data-syndrome codes \cite{ashikhmin2020quantum} for example. 
    \subsubsection{Surface and Color Codes}
    Our compiler is tailored towards optimizing the broader class of stabilizer codes, but decoding each stabilizer code efficiently is nontrivial. For this reason, we evaluate our compiler's effects on the logical error rate for two of the most-studied and promising class of quantum error-correcting codes, surface and color codes, as representative samples \cite{kitaev1997toriccodes, dennis2001surfacecodes, landahl2011colorcodes, tomita2014surfacecodewithrealisticnoise, lee2024concatmwpm}. Figure \ref{fig:backgroundcodes} shows the high-level structure of surface and color codes, respectively. On both code lattices, the vertices of the codes correspond to data qubits. For faces, the structure is different. For surface codes, each face corresponds to either an $X$ or $Z$ stabilizer (one for each color blue or green). In contrast, for color codes, each face corresponds to \textit{both} an $X$ and $Z$ stabilizers.
    
    % Table~\ref{tab:code_properties} contains other pertinent information about both codes, namely (1)~the number of data qubits in a distance $d$ logical qubit, (2)~the number of stabilizers in a distance $d$ logical qubit\footnote{Note that the stabilizers are split evenly between $X$ and $Z$ for both codes.}, (3)~the check weight, or number of data qubits in a stabilizer, for each code, and (4)~the decoding algorithms used for each code.

    % \begin{table}[!htb]
    %     \centering
    %     \begin{center}
    %         \caption{Properties of Surface and Color Codes}
    %         \label{tab:code_properties}
    %         \begin{tabular}{|c||c|c|}
    %             \hline
    %             & Surface Code & Color Code \\
    %             \hline
    %             \hline
    %             Data Qubits ($n$) & $d^2$ & $\frac{3}{4}(d^2-1)+1$ \\
    %             \hline
    %             Stabilizers & \multicolumn{2}{c|}{$n-1$} \\
    %             \hline
    %             Maximum Check Weight & $4$ & $6$ \\
    %             \hline
    %             Decoder & PyMatching~\cite{higgott2023sparseblossom} & Chromobius~\cite{gidney2023chromobius} \\
    %             \hline
    %         \end{tabular}
    %     \end{center}
    % \end{table}

\subsection{QCCD Trapped Ion Systems}
\begin{figure}
    \centering
    \includegraphics[width=\linewidth]{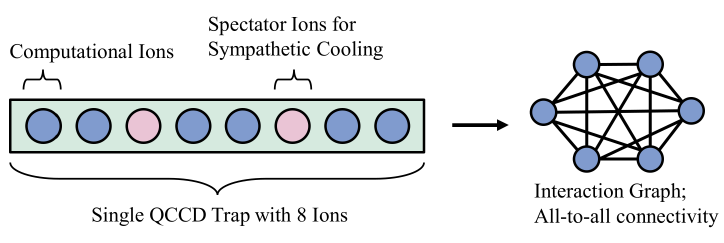}
    \caption{QCCD trap abstraction. Traps contain a combination of computational ions, to which data and ancilla qubits can be mapped to, and spectator ions used to keep the other ions cool. As gates are executed and ions moved, heat accumulates typically resulting in lower fidelity operations making cooling ions necessary. We treat cooling as a background process. The remaining computational ions can be interacted in any combination but only one operation per trap per time step.}
    \label{fig:qccd1}
\end{figure}
\begin{figure}
    \centering
    \includegraphics[width=\linewidth]{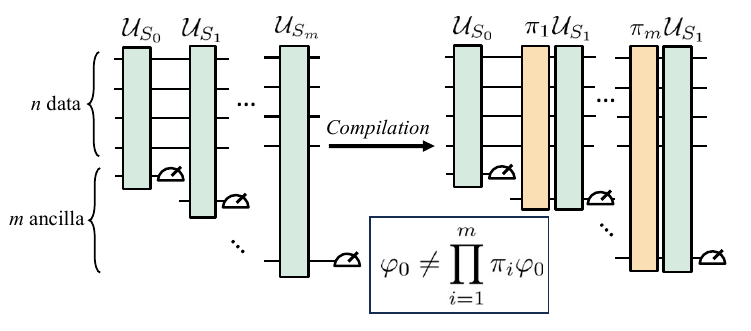}
    \caption{Compilation for these programs transforms the sequence of stabilizer circuits $\mathcal{S}_i$ on $n$ data and $m$ ancilla (left) into one which is compliant with the restrictions of the target QCCD device (right). Qubits are mapped into each trap given by $\varphi_0$. Because shuttling must be added, this mapping gets permuted over time. Here we represent it as $\pi_i$. A NISQ compiler which compiles a single round of extraction is unaware of repeated execution and so the end of this circuit has a mapping $\prod_{i = 1}^{m} \pi_i\varphi_0$ which is likely different than the initial mapping, meaning an expensive step must be inserted to repeat this circuit.}
    \label{fig:latencyexplanatory}
\end{figure}

\subsubsection{Limitations on Connectivity}
\begin{figure*}
    \centering
    \includegraphics[width=\linewidth]{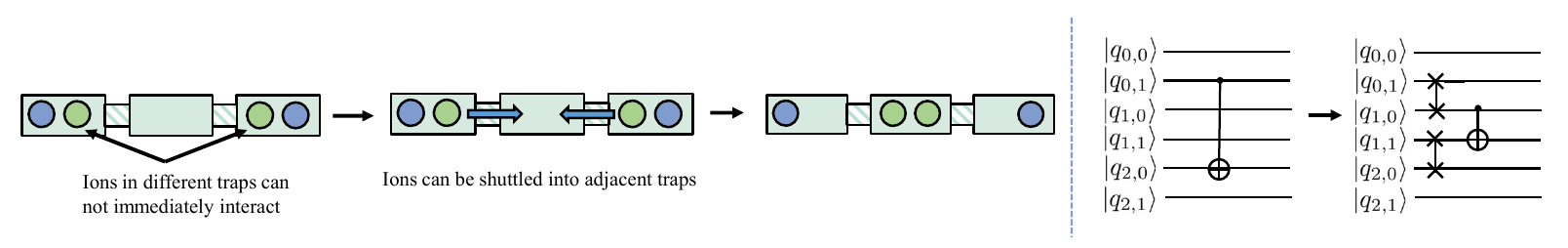}
    \caption{QCCD movement overhead is induced by low intra-trap connectivity. Long range gates on qubits (e.g. the two green colored circles here) in different traps require some shuttling overhead to relocate one (or both) qubits into the same trap prior to gate execution. Shuttling consists of a multi-stage sequence of splits and merges but can more simply be represented as a SWAP between traps in the circuit model. We account for all associated errors when simulating, however.}
    \label{fig:qccd2}
\end{figure*}

Despite their high fidelity and high coherence times, the fundamental limitation of scaling trapped ion systems lies in increasing the number of physical qubits that can be controlled concurrently. Given this limitation, the \textit{Quantum Charge Coupled Devices (QCCD)} architecture has emerged as a promising method of scaling trapped ion quantum computers. QCCD architectures are modular, as they contain multiple functionally identical traps, and within each trap, all-to-all connectivity is supported as seen in Figure \ref{fig:qccd1}. To entangle qubits within different traps, a qubit from one trap must be shuttled into another trap via an interconnect, for example in Figure \ref{fig:qccd2}.

\subsubsection{Limitations on Parallelism}

Nevertheless, despite having all-to-all intra-trap connectivity, QCCD architectures will likely only support one or two concurrent two-qubit operations within a single trap due to the calibration overheads associated with supporting concurrent two-qubit operations\footnote{In this paper, we assume that each trap in a QCCD architecture can only perform \textit{one} two-qubit operation.}. Consequently, QCCD architectures exhibit low intra-trap parallelism. Operations in \textit{different} traps and shuttling operations are unconstrained in terms of parallelism; however, for near term devices with low amounts of traps, operations across the entire system are nearly serial.  

\subsection{Shuttling Overheads}

Since QCCD hardware provides all-all connectivity inside traps but not between traps, ions must be moved or ``shuttled'' from a source trap to a shared destination trap to enable full device connectivity. The shuttling process can be broken up into different phases: splitting, swapping, moving, and merging. Each of these processes contributes to overall heating, and adds time overhead; both contribute to higher logical error rate.

To shuttle, one or more ion(s) are split from the end of a chain, and placed into a shuttling zone with a total cost of 80 microseconds. Moving through the shuttling zone takes 5 microseconds, unless encountering a merge of shuttling zones in an X or Y junction to where an additional 120 or 100 microseconds are used (respectively). Finally, the ion(s) are merged into the destination trap. There are two methods to move ions inside the trap: GateSWAP and IonSWAP. GateSWAP implements a SWAP gate (3 CX) between the source and target ion and therefore takes $3g = 300$ microseconds, where $g$ is the gate time. IonSWAP on the other hand physically rearranges the trap to swap the ions, and takes time dependent on the number of ions in the trap (e.g. $= 42n$) where $n$ is the number of ions \cite{ionswap}. Just as gate operations can be parallelized across different traps, shuttling operations across different traps can also be parallelized. 

When compiling a circuit to QCCD hardware, shuttling operations are interleaved with gate operations as shown in Figure \ref{fig:latencyexplanatory} with the example of a QEC code. Note that for any quantum circuit that must be run across many cycles (like QEC codes), the mapping and ensuing connectivity is unlikely to be the same at the end of the cycle as it was at the beginning of the cycle, presenting a challenge for current QCCD compilers to compile thousands of rounds of syndrome extraction circuits.

\begin{figure}
    \centering
    
   \includegraphics[width=200pt]{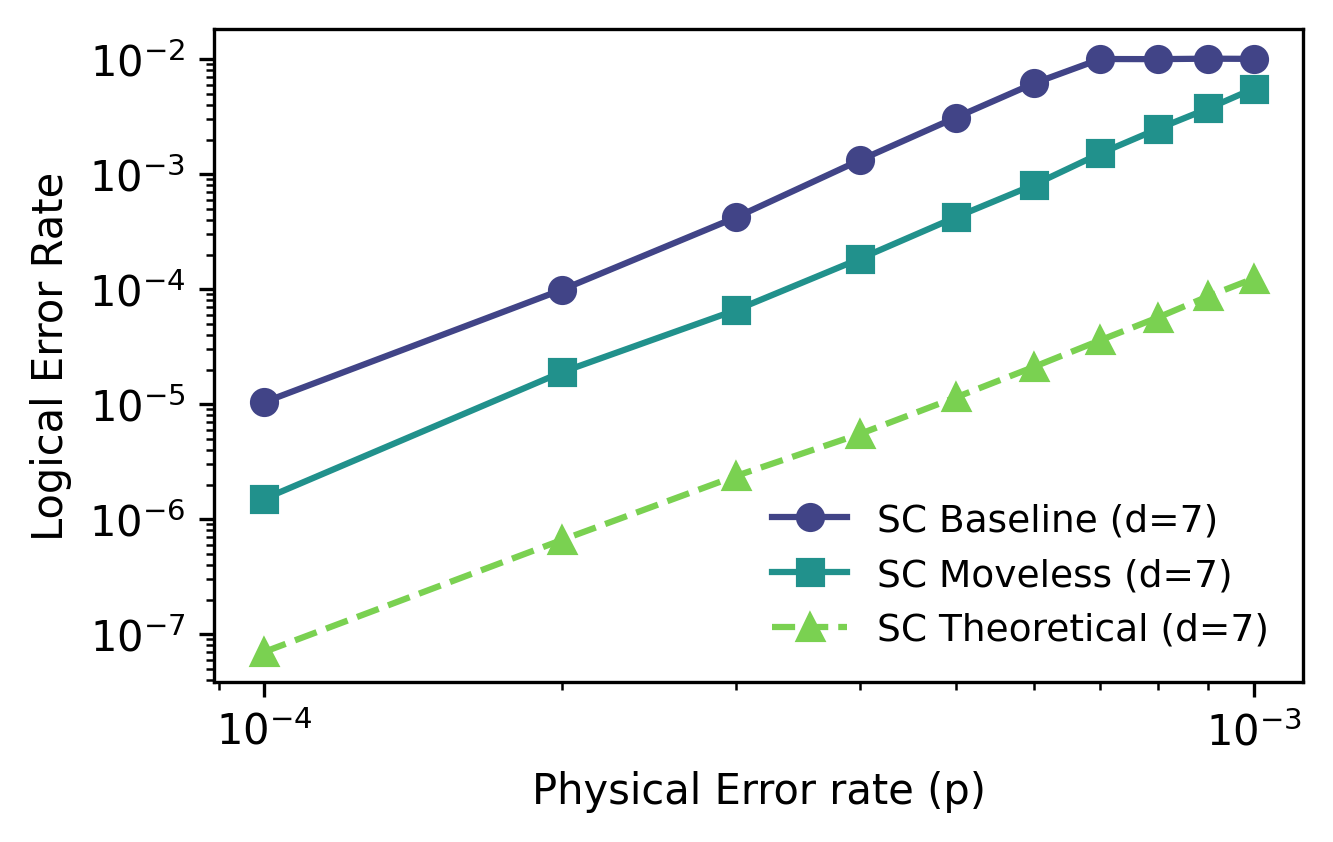}
   \caption{In the case of zero hardware constraints (shuttling is free, maximum parallelism) the \textit{theoretical} distance 7 surface code latency is extremely small. Our work bridges this gap purely through software means using efficient compilation for QEC}
    \label{fig:BackgroundDepthAnalysis}
\end{figure}
\subsection{Syndrome Extraction Circuits}
Syndrome extraction circuits, which entangle data qubits with ancilla qubits to measure the stabilizers of a quantum code, are responsible for detecting errors and forming a syndrome to be used by a decoder. Consequently, the success of quantum error correction depends on the fidelity of a syndrome extraction circuit. 

There are two factors that determine syndrome extraction fidelity: (1)~operation error, and (2)~latency. In this paper, we focus on reducing the execution latency of syndrome extraction circuits on QCCD trapped ion systems, as the gate and measurement operations performed \textit{on a given code} are identical regardless of compiler or architecture. As syndrome extraction is a periodic process, between syndrome extraction rounds, data qubits can incur errors due to dephasing or heating. These effects worsen with longer syndrome extraction circuits, as errors on the data qubits will remain undetected for longer periods of time. On QCCD systems, there are two barriers towards achieving short syndrome extraction circuits: (1)~limited intra-trap parallelism, which effectively serializes most quantum operations, and (2)~expensive shuttling overheads, as each trap can only house 5-10 physical qubits at once.

\subsection{Prior Work and Their Limitations}
Much like any other quantum circuit, syndrome extraction circuits also must be compiled to the QCCD architecture. To the best of our knowledge, the only available compilers are \textit{Noisy Intermediate Scale Quantum (NISQ)} compilers, which are optimized for running \textit{general} quantum circuits on physical hardware. As these compilers are not optimized for syndrome extraction, the heuristics leveraged by these compilers often make incorrect shuttling decisions during compilation, resulting in high syndrome extraction overhead.

Figure \ref{fig:BackgroundDepthAnalysis} demonstrates the effects of deep syndrome extraction circuits on the logical error rate of a $d = 7$ surface code at varying physical error rates. Assuming no architectural constraints, implying the \textit{theoretical} amount of parallelism and connectivity could be achieved, the total execution time is bound to $4 * CX + 2*H + M$ where CX, H, and M are the two qubit gate latency, single qubit gate latency, and measurement latency respectively. As shown, the gap is large - spanning orders of magnitude in logical error for just a medium-sized QEC code. Our motivation in this work is to bridge this gap purely by using efficient software compilation, and even for a medium size code we show how this translates to around an order of magnitude improvement of logical error rate.

% \cite{tomita2014surfacecodewithrealisticnoise}. As shown, a $10\times$ increase in the circuit depth results in a small increase in the logical error rate. For comparison, a near $50\times$ increase in depth  results in roughly an order of magnitude increase in the logical error rate, close to disallowing practical error correction as the logical error rate is roughly equal to the physical error rate. Optimizing the depth of syndrome extraction circuits is necessary for practical implementations of quantum error correction, especially in the near-term when error rates are higher.

% Futhermore, Figure \ref{fig:BackgroundDepthAnalysis} also shows that on a $d = 7$ surface code, a baseline compiler \cite{murali1} yields a syndrome extraction circuit that runs for around 0.1 seconds, which is around 236$\times$ longer than an idealized, fully parallelized syndrome extraction circuit, demonstrating the sizable room for improvement. In fact, at $p = 10^{-3}$, its syndrome extraction circuit is unusable for error correction, as the logical error rate is \textit{worse} than the physical error rate.

% \subsection{Goal}
%     Reducing syndrome extraction depth is necessary for enabling error correction on QCCD systems. In this paper, we tackle the problem of reducing syndrome extraction circuit depth on QCCD architectures.

\section{Understanding Limitations of Existing Compilers}
\label{sec:limitations_of_compilers}
%existing compilers NISQ
%Compilation pipeline involves mapping
\begin{figure*}
    \centering
    \includegraphics[width=\linewidth]{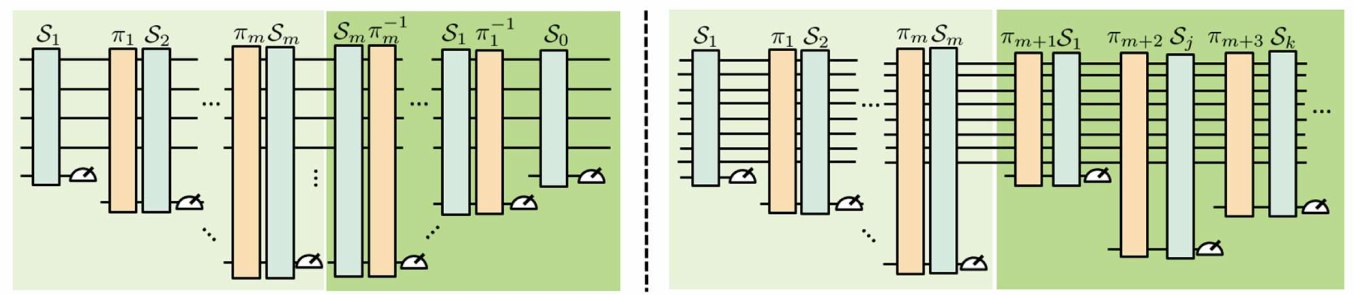}
    \caption{On the left, we find a movement pattern $\pi_1, ..., \pi_m$ while performing stabilizers $S_1, ...., S_m$. To proceed with executing the circuit across multiple cycles, we pair the circuit with its inverse order $S_m, ..., S_1$ and $\pi_m^{-1}, ..., \pi_1^{-1}$. On the right we show the process without the inverse mapping: possibly an entirely new order of movement patterns and stabilizers.}
    \label{fig:reversecircuit}
\end{figure*}
Syndrome extraction has additional latency on QCCD hardware due to the insertion of excessive shuttling overhead (e.g. in Figure \ref{fig:latencyexplanatory}). The compilation pipeline for syndrome extraction circuits on QCCD devices includes mapping and scheduling, i.e. assigning program qubits a hardware location, inserting additional movement operations for hardware qubits (shuttling), and ordering the gate operations. Optimally solving each of these problems is challenging, and most solutions resort to heuristics that optimize for \textit{NISQ} based program execution. This is the case for prior work in QCCD architectures \cite{murali1, Saki_2022}. Because compilation is \textit{NISQ} focused, these heuristics are designed to be general and adaptive. While flexibility is useful for enabling users to submit arbitrary programs that are assumed to have little to no structural regularity, it is extremely inefficient for syndrome extraction circuits that have such structural regularity. This additional overhead on syndrome extraction translates to more damping and dephasing error and ultimately, an increased logical error rate.
\begin{figure}
    \scalebox{.8}{
   \includegraphics[width=\linewidth]{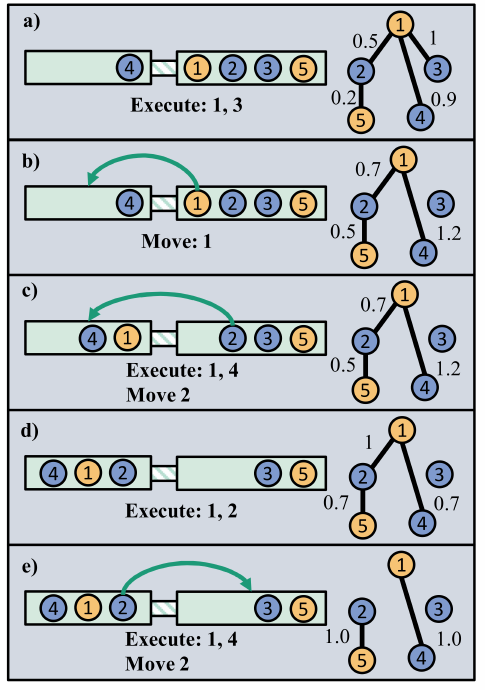}}
    \caption{Shuttling operations are shown through 5 different time steps for a baseline compiler that moves both ancilla and data. The interaction graph is shown on the right of the trap, where higher edge weights indicate a better heuristic score for scheduling. In this case, qubit 2 was unnecessarily ``pulled" between traps, when shuttling only qubits 1 and 5 once suffices.}
    \label{fig:ancillamovement}
\end{figure}
\subsection{Case Study: Sample Execution}
\subsubsection{Unconstrained Qubit Movement}
In Figure \ref{fig:ancillamovement}, we explore some of the pitfalls of these generic compilation methods. Here, we construct a simplified version of a typical QEC circuit. These circuits can be represented as a bipartite Tanner graph between ancilla (yellow) and data (blue) \cite{leverrier2022quantumtannercodes}. The interaction graph can be used to determine which operation to perform next, e.g. using a lookahead function \cite{10.1145/3387902.3392617}; here the larger the value the higher priority for immediate execution, with a value $\ge 1$ indicating there is a blocking interaction between the given pair. These operations must be executed before the circuit can continue and will force ion rearrangement if not located in the same trap. Furthermore, these compilers must be given a finite circuit, e.g. a single round of syndrome extraction or a small batch. As the circuit executes, the graph will become disconnected, resulting in data and ancilla moving more freely through different traps. These heuristics result in data and ancilla ``pulling'' each other between traps leading to excessive movement overheads and similarly will result in an arrangement of the ions which is not necessarily good for the next round of extraction, and must then be recompiled for a new starting position. 

\subsubsection{Limited Circuit Input}
NISQ compilers used out of the box will compile based on a single or a handful of syndrome extraction rounds. For example, in Figure \ref{fig:latencyexplanatory} the qubit mapping permutations introduced to execute each stabilizer measurement will result in a qubit mapping after execution which is likely drastically different than the ``optimal'' initial mapping. However, because syndrome extraction will be executed repeatedly for hundreds of thousands of rounds, this leads to excessive movement to rearrange between rounds if repeated naively. A simple alternative is given in the left of Figure \ref{fig:reversecircuit}. Given an optimal (or near optimal) movement pattern, $\pi_1, ..., \pi_m$, to execute each of the stabilizers $S_1, ..., S_m$, a subsequent measurement of the stabilizers in the reverse order, $S_m, ..., S_1$, can be achieved by simply inverting the movement patterns from the first cycle, $\pi_m^{-1}, ..., \pi_1^{-1}$ and then repeated again. Then after every two cycles, we return to the initial mapping allowing us to execute the same circuit pattern again. While simple, this strategy is effective and enables out of the box compilers to be used for these infinitely repeating circuit patterns. 

\begin{figure*}
   \includegraphics[width=\linewidth]{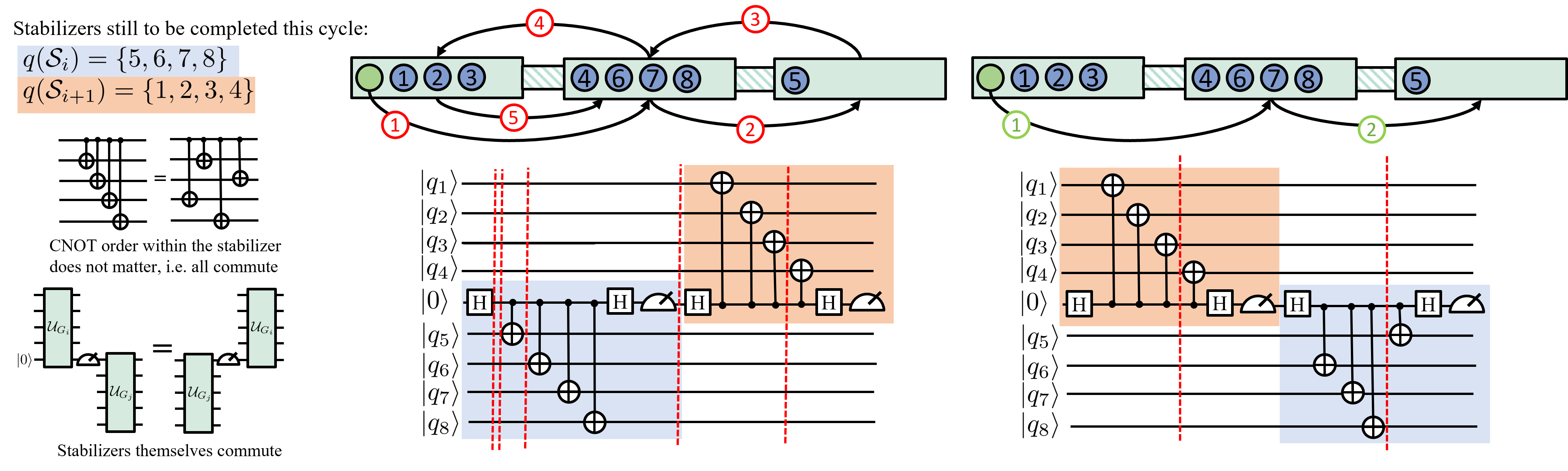}
    \caption{On the left, the two key principles of the Moveless dynamic scheduling algorithm are shown: stabilizers commute each other and CNOT order within stabilizers can be fully relaxed. In the middle, the static ordering is shown, and each vertical line in red denotes a corresponding shuttling operation inserted, for a total of 5. On the right, dynamic scheduling is shown, reordering the orange $S_{i+1}$ first and executing these pair of stabilizers in only 2 shuttling operations.}
    \label{fig:movelessFigX}
\end{figure*}

% \begin{figure}
%    \includegraphics[width=\linewidth]{figures/moveless-pt3.png}
%     \caption{Moveless's dynamic scheduling algorithm can be improved with ancilla reuse. Originally ion A (purple) is assigned to 2 stabilizers and B (red) 1 stabilizer off a naive predetermined ordering (top). Dynamically assigning stabilizers allows for efficient reuse (bottom).}
%     \label{fig:movelessFigY}
% \end{figure}

\begin{figure*}
   \includegraphics[width=\linewidth]{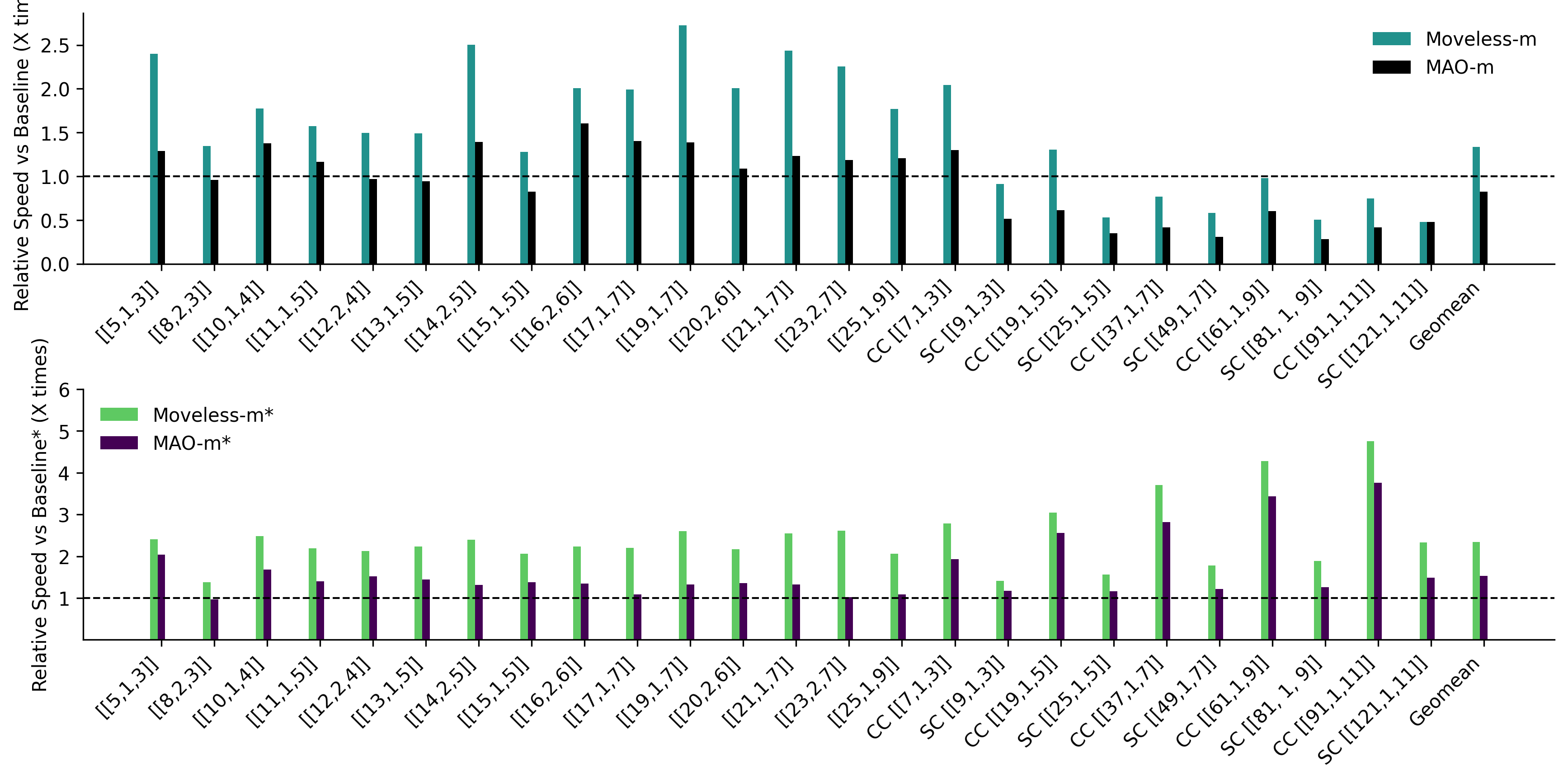}
   \caption{Execution time results for a suite of arbitrary Stabilizer Codes and then Color Codes (CC) and Surface Codes (SC). We compare using the factor of relative improvement over the Baseline (top) and Baseline* (bottom), such that 1 = the execution time of the Baseline/Baseline*. On the shorter end, execution times are in the mere single thousands of microseconds such as the distance 3 color code. On the longer side, execution times are on the order of seconds such as the distance 11 color code.}
    \label{fig:stabilizer_barchart}
\end{figure*}

% \begin{figure*}
%    \includegraphics[width=400pt]{figures/MMOvsBaselinePlots.pdf}
%     \caption{Threshold plots between MMO and baseline compilers on the surface and color code. The threshold is marked with a vertical dashed line.}
%     \label{fig:thresholdhighbaseline}
% \end{figure*}
% \begin{figure}[h]
%      \centering
%      \begin{subfigure}[b]{200pt}
%          \centering
         
%          \includegraphics[width=\textwidth]{figures/SCAncillaSensitivity.pdf}
%          \caption{Distance 7 Surface Code}
%          \label{fig:SCancillanalysis}
%      \end{subfigure}
%      \begin{subfigure}[b]{200pt}
%          \centering
         
%          \includegraphics[width=\textwidth]{figures/CCAncillaSensitivity.pdf}
%          \caption{Distance 7 Color Code}
%          \label{fig:CCancillanalysis}
%      \end{subfigure}
%     \centering
%     \caption{{\color{red}We study the effect of changing the amount of ancilla (A) relative to the max ancilla (MA) on both the distance 7 surface code (top) and color codes (bottom). We find the general trend that reusing ancilla tends to decrease logical error rate for our MMO compiler.}
%      \label{fig:ancillanalysis}}
% \end{figure}
\section{Minimizing Intertrap Movement and Adaptive Scheduling for QEC}

We seek to reduce the excessive syndrome extraction depth on QEC circuits through the reduction of total shuttling operations and an adaptive ordering of stabilizers in a compiler we ultimately designate as Moveless.

\subsection{MAO: Moving Ancilla Only}
% \textbf{FIRST TALK ABOUT HOW CURRENT COMPILERS MAP, REBALANCE, TRANSITION INTO }
We aim to minimize the need for intertrap movements, thus minimizing total shuttling overhead and reducing syndrome extraction latency. During compilation, qubits are first mapped onto underlying hardware with the constraint that each machine has a trap capacity. Traps are packed moderately dense as to be spatially efficient and retain higher qubit connectivity but not become overcrowded. If a trap is too overcrowded, \textit{rebalancing} is needed to spread qubits more evenly across each trap to prevent a bottleneck in new qubits being moved to and from the overcrowded trap.

After mapping, the minimization of movement involves minimizing two qubit interactions in different traps. We exploit the structural regularity derived from the graph representation of a syndrome extraction circuit and use this knowledge to create policies that will minimize overall intertrap interactions. We know that each syndrome extraction circuit could be represented as a bipartite interaction graph in the form of a Tanner graph with an ancilla and data partition \cite{leverrier2022quantumtannercodes}. This implies any syndrome extraction circuit can be completed without ancilla ever needing to interact with other ancilla, and data needing to interact with other data. Meaning that if all the data in are distributed over $n$ traps, then we can ensure every necessary interaction is enabled by moving \textit{only} ancilla between traps (conversely, data can be distributed across ancilla distributed in $n$ traps). In fact, we observe each ancilla needs to, in the worst case, enter and exit each trap exactly once (per stabilizer) if the data is not allowed to move. To bound the movement lower than $n$, we can pack data as tightly as possible. Therefore, if we move only the ancilla between partitions of data across all traps, the total possible movement is heuristically minimized while still ensuring the correctness of circuit execution, remedying the ``pulling" problem shown in Figure \ref{fig:ancillamovement}.

We implement this heuristic policy in a compiler we designate as \textit{Move Ancilla Only (MAO)}. We partition data qubits into traps based on prior graph fine grained partitioning work \cite{10.1145/3387902.3392617}, with the aim of creating partitions that minimize total movement between traps. The partitioner divides the circuit into time slices at each layer and considers both future interactions and nearby interactions at each time slice in order to find mappings that heuristically have a high amount of interactions per trap. We are then able to move only ancilla between these partitions of data.

\subsubsection{Analysis of Initial Improvement for MAO}

The MAO heuristic does moderately well across our initial suite of stabilizer codes as shown in the top row of Figure \ref{fig:stabilizer_barchart}, in general exhibiting modest speedup. However, the struggle to utilize multiple ancilla  
efficiently on denser partitions of data lead to losses against the baseline in larger, more parallelizable codes like surface and color codes. 
\subsubsection{Roadblocking While Scaling to Larger Codes}
As seen in Figure \ref{fig:stabilizer_barchart}, upon moving to larger and more parallelizeable codes, only being able to move ancilla becomes somewhat costly when there are lots of ancilla and low relative opportunity for parallelism due to greater ``roadblocking" potential in a linear architecture (where a larger amount of near full traps don't have flexibility to move data, triggering rebalances often as multiple ancilla enter), suggesting that MAO in its base form needs modification. We later address this in Section \ref{subsec:ancilla_reuse} by updating our compiler (MAO*) to use ancilla more efficiently. We further show in architectural sensitivity experiments (Figure \ref{fig:ancilla_grid_amounts}) how this value of ancilla changes on grid architectures when roadblocks aren't as easy to create. 

\subsection{Moveless: Dynamic Stabilizer Scheduling + MAO}

We build upon MAO's movement minimization policy and add a QEC appropriate scheduling mechanism. We further leverage the idea detailed in Section \ref{sec:limitations_of_compilers} of reordering stabilizers to dynamically find a good ordering for the set $\{S_1 ... S_m\}$ according to the data/ancilla arrangement at each scheduling interval, instead of assuming the static ascending ordering. We break up the scheduling process into \textit{intervals at the stabilizer level} due to constraints stabilizer codes have in parallelizing execution of multiple stabilizers at once \cite{gottesman1997stabilizer}. Instead of viewing the circuit as a DAG of gate dependencies as baseline compilers do, we scrap this rigid representation in favor of a set representation of stabilizers and a scheduling-queue we iteratively add to at each time step. Within each time step, we use a greedy algorithm to compute movement scores of all possible candidate stabilizer-ancilla pairs based on a shortest path algorithm between traps on the hardware's connectivity graph. The stabilizer-ancilla pair that achieves the minimum cumulative score for all gates throughout the pair's execution is then used for its gates to be added to the scheduling queue. All ancilla-stabilizer pairs are eligible, including ancilla that were already used. Once the stabilizer/ancilla pair is chosen, the order of each gate within the stabilizer is chosen so that the gates with the lowest individual movement score are chosen first. After the gates are efficiently ordered within the stabilizer, they are added to the scheduling queue, and the stabilizer is removed from the stabilizer set, where this process repeats until the stabilizer set is empty and the scheduling queue is full. These gates are then broke into atomic steps: split, merge, move, and true gate execution time. To parallelize operations, an analyzer iterates through the scheduling queue and parallelizes the atomic operation if no other trap resources are being used.  Figure \ref{fig:movelessFigX} demonstrates this process in detail: At the time step shown, the orange $S_{i+1}$ stabilizer is more optimal than performing the initially scheduled blue $S_i$ (accounting for the fact that gates can be reordered within the stabilizer, but in this case it makes no difference). It then performs the orange stabilizer, and then repeats the process of scheduling the next best stabilizer (again, accounting for relaxed order between gates of stabilizers). This process repeats throughout the execution of the entire circuit.

\subsection{Reusing Ancilla}
\label{subsec:ancilla_reuse}
In accordance with our selection of minimal movement heuristics, we leverage the idea that since ancilla can be reused for different parity checks upon completion of another check, keeping $m$ ancilla in the circuit (where $m$ denotes the number of stabilizers) may be harmful to execution time and cause more overall intertrap movement because they may be assigned to perform a measurement but idle for long periods of time before actually used. We experiment with different amount of ancilla on all three compilers (Baseline, MAO, and Moveless) and take note of the varying syndrome extraction latency as shown in the bottom row of Figure \ref{fig:stabilizer_barchart}. We find that lower amounts of ancilla generally lowers shuttling overhead, suggesting there is some optimal value $m^* \le m$. We compare six different values of ancilla counts across all of our codes: $1, 20\%m, 40\%m, 60\%m, 80\%m, m$. We find that for linear architectures, the extreme case where there is only one ancilla, $m^*$ = 1, works best for both MAO and Moveless across our entire suite of codes (but revisit this assumption when the architecture is changed from a linear architecture in Figure \ref{fig:ancilla_grid_amounts}). For the baseline, this value is not always 1 but is generally lower than $m$ ancilla qubits. We give the baseline the optimistic assumption that it can pick the best of these values in Baseline$*$ as shown in the bottom row of Figure \ref{fig:stabilizer_barchart}. We postulate that the compilers do better with lower amounts of ancilla than $m$ for 3 reasons: 1) Due to hardware constraints on low parallelism and theoretical constraints on parallelization between stabilizers, multiple ancilla are difficult to utilize effectively 2) ancilla create roadblocks especially within linear architectures, triggering rebalances often and saturating trap capacity 3) In the specific case of Moveless, a particular ancilla could be remapped to many stabilizers, encouraging reuse. 

In the bottom row of Figure \ref{fig:stabilizer_barchart}, we find that Moveless (now updated with full ancilla reuse) unanimously outperforms MAO$^*$, which unanimously outperforms the baseline with its respective optimal ancilla value across all six choices. Moveless has a best case speedup of $5.24\times$, worst case speedup of $1.4\times$, and the geometrical mean of speedups is of $2.34\times$ across our entire suite of codes.

\section{Evaluation Methodology}
 \begin{figure*}
    \centering
    \includegraphics[width=\linewidth]{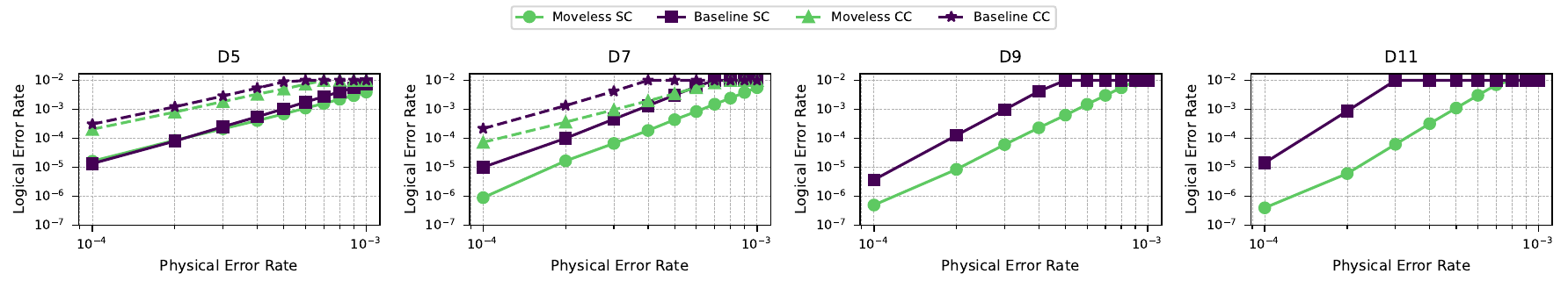}
    \caption{We compare the logical error rate difference caused by the additional amplitude damping/dephasing error injected due to long syndrome extraction latency. We compared color codes (CC) and surface codes (SC) for distances 5-11}
    \label{LER:SCandCC}
\end{figure*}
 We build our QCCD compilers on top of the environment offered by \cite{murali1} which simulates the costs of movement in QCCD in addition to their compiler. We perform our experiments on a suite of generalized stabilizer codes and collect syndrome extraction latencies from this framework \cite{murali1}. We then perform a realistic simulation of the effect this latency has on logical error rate for surface and color codes between $d=5$ and $d=11$ as realistic distances which could be implemented on soon-to-be-realized hardware. We expect these codes to be used on near-term trapped ion systems within the next decade.

\subsection{Architectural Setup}
\subsubsection{Hardware Topology}

We evaluate a total of 3 compilers throughout this paper (Baseline \cite{murali1}, MAO, and Moveless) on the same linear connection topology as a reasonable and simple experimental constant. This linear configuration is of $m$ traps where $m$ is the number of parity checks within a given code, therefore allowing any of the compilers the ability to potentially utilize as much parallelism as the code theoretically could allow if all its stabilizers were independent (though a key result of our paper is that this amount is still very low). We later experiment with how our findings change with differing connection topologies, like a grid in Section \ref{sec:grid_sensitivity} 
\subsubsection{Experimental Constants} In our simulations, there are many critical design choices we hold constant. We hold a constant ion trap capacity of 5 ions per trap, which is a typical amount in QCCD traps and found in prior work to be a reasonable choice \cite{ovide2024scalingassigningresourcesion, murali1}. We use the GateSWAP technique for intratrap ion movement, gate times obtained from phase modulated gates as a function of ion trap capacity, and shuttling times as found in \cite{murali1}. Since many of these assumptions have a significant effect on experimental outcome, we later explore sensitivity to some of these parameters as well as comparing against other baseline topologies that are not linear. Since a syndrome extraction circuit has no implicit ordering on its stabilizers, we assume MAO and the baseline are able to perform their reverse circuit on even rounds as shown in Figure \ref{fig:reversecircuit}.

\subsection{Noise Model and Evaluation of Logical Error Rate}

\subsubsection{Core Error Model} We consider physical error rates $p$ for values around the threshold for surface and color codes ($10^{-3}$ and $10^{-4}$). We inject circuit-level noise at $0.1p$ for single-qubit gate error, and at $p$ for two-qubit gates and measurement errors. We use the PyMatching \cite{higgott2022pymatching} and Chromobius \cite{gidney2023chromobius} decoders on Surface Codes and Color Codes, respectively, in order to decode syndromes. We combine this core error model with the additional error due to syndrome extraction latency and sympathetic cooling latency to create a custom noise model that more accurately simulates the performance of quantum error correcting codes on trapped ion systems.

\subsubsection{Modeling Syndrome Extraction Latency as a Depolarizing Channel}
For each of the 3 designs, we compute the syndrome extraction latency for each code by finding their shuttling and gate times through compilation. In order to accurately model this latency as a depolarizing channel in terms of time-dependent amplitude damping and dephasing errors (associated with T1 and T2 times), we use the Pauli twirling approximation~\cite{tomita2014surfacecodewithrealisticnoise}. We inject these errors on each qubit after each round of syndrome extraction. We bound T1 and T2 times between 10-100s as reasonable estimates from available hardware\cite{progress_in_QC}, and use a logarithmic fit for these T1 and T2 times to change with $p$ (where $p = 10^{-3}$ corresponds to T1 and T2 $= 10$ seconds and $p = 10^{-4}$ corresponds to T1 and T2 $= 100$ seconds).

\subsubsection{Sympathetic Cooling}
In order to avoid accumulating heating errors throughout circuit execution, we assume sympathetic cooling  after each merge operation. Two extra ions are added into each trap, and when laser-cooled, they reduce the overall heat from the system. As detailed in \cite{sympathetic-cooling}, this can be implemented to practically eliminate the effects of heating if done after every operation. However, this does incur additional time which is added to the entire syndrome extraction round's total latency.

\section{Final Evaluations}
\subsection{Analysis of Logical Error Rates}

\subsubsection{Improvement of Logical Error Rate}
    We model the reduction of syndrome extraction depth and its reduction in amplitude damping and dephasing errors, ultimately leading to lower logical error rate on surface and color codes given larger execution times. Our Moveless compiler unanimously has shorter execution times compared to the baseline and therefore exhibits logical error rate improvements (or matched performance) across the board for both surface/color codes for all values of $p$ and $d$. We find that the magnitude of our improvement in logical error scales with the code size, as the distance 11 surface code demonstrates an improvement of logical error rate from $10^{-5}$ to $10^{-7}$ as shown in Figure \ref{LER:SCandCC}. 

\subsubsection{Threshold ``Slipping"}
A code threshold is traditionally referred to as the physical error rate $p$ where increasing code distances guarantees a reduction in logical error rate. Thresholds are typically seen as a property of a code when assuming certain standardized error models. However, when mapped to trapped ions, we have detailed a more complex and custom noise model for these codes that introduce additional scaling challenges (i.e. larger codes require more gate and shuttling times and thus more amplitude damping and dephasing, which themselves are parameterized from a log-fit of $p$). Because of this, we observe threshold ``slipping" in our evaluations\cite{svore2006flowmapmodelanalyzingpseudothresholds}\cite{Pseudothresholds1}: where scaling to a higher distance may necessitate a lower $p$ to see improvements in logical error rate. For this reason, instead of comparing values of a threshold for a given code family, we observe differences in logical error rate at given values of p. This is standard when comparing practical implementations of codes as opposed to purely theoretical models.

\subsection{Sensitivity Analysis}
\label{sec:grid_sensitivity}
\subsubsection{Sensitivity to Architecture}
\begin{figure}
    \centering
    
   \includegraphics[width=200pt]{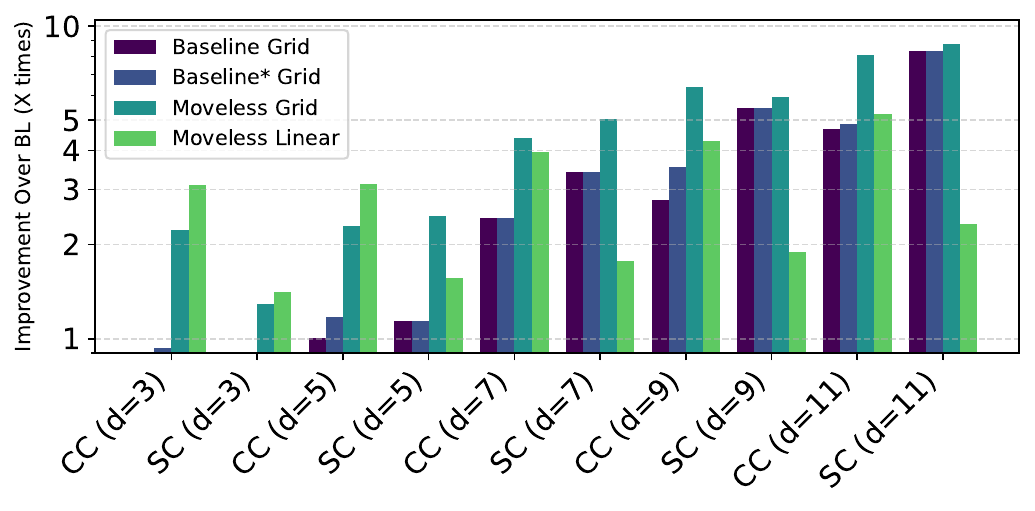}
   \caption{We analyze the difference between a maximally parallel grid architecture vs our original maximally parallel linear architecture. We compare fractional improvements relative to the linear baseline (BL). Experiments were run with 60 traps (again allowing for large amounts of theoretical parallelism) in a 6x10 mesh.}
    \label{fig:gridSensitivity}
\end{figure}

\begin{figure}
    \centering
    
   \includegraphics[width=200pt]{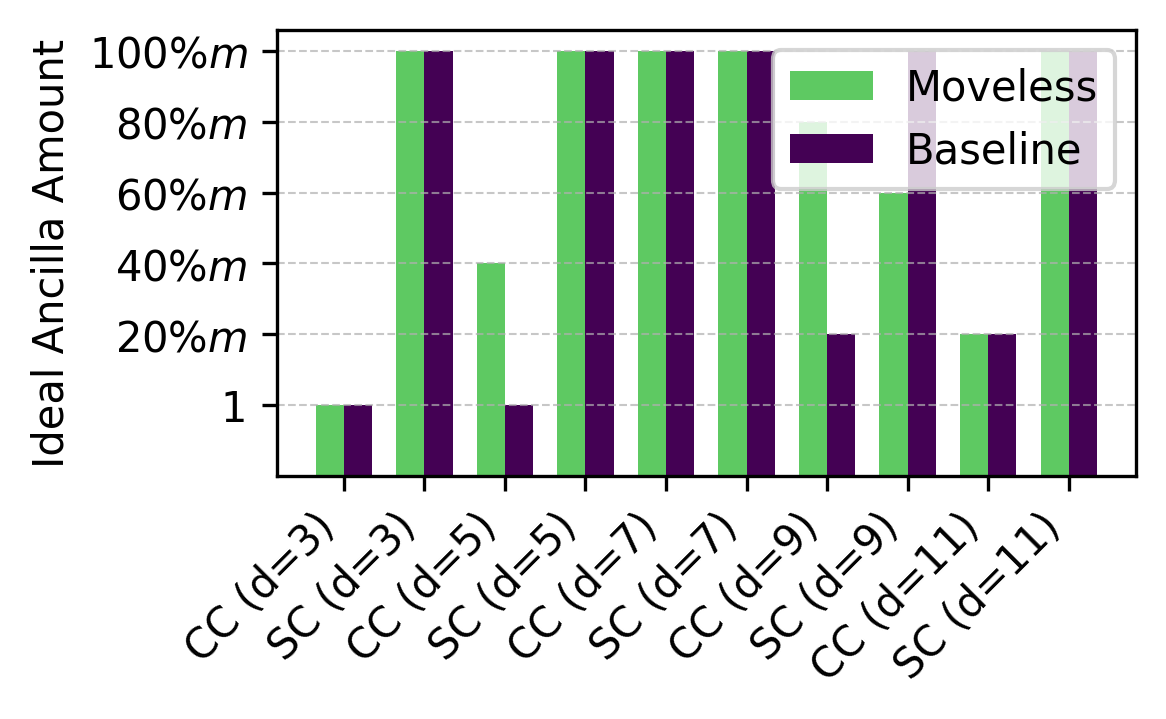}
   \caption{We can utilize a linear on $m$ ancilla search between values of ancilla count in the circuit to find the best one for Moveless on the Grid architecture. For the baseline, this value also fluctuates. In this plot, 1 indicates the extreme case of only 1 ancilla, while the others indicate a percentage of the maximum ancilla $m$}
    \label{fig:ancilla_grid_amounts}
\end{figure}

Since the architectural design space for QCCDs is still being explored, we model our compilers' benefits on other hardware connection topologies. In Figure \ref{fig:gridSensitivity}, we still find unanimous wins across the board for Moveless. Grid architectures have shuttling junctions which allow for 2D topologies at the expense of additional shuttling time. For this experiment, we take junction cross times as found in \cite{murali1}. For this reason, smaller size codes perform worse on grid topologies, but as code size increases the grid architecture becomes more valuable. Due to the much lower susceptibility to ``roadblocking" as described in previous sections, the grid architecture allows for more parallelism and more ancilla usage. Interestingly, we find that this changes the optimal value for $m*$ for Moveless, and explore the differences in this choice in Figure \ref{fig:ancilla_grid_amounts}.  

\subsubsection{Sensitivity to Ion Capacities/Trap Counts}

\begin{figure}
    \centering
    
   \includegraphics[width=200pt]{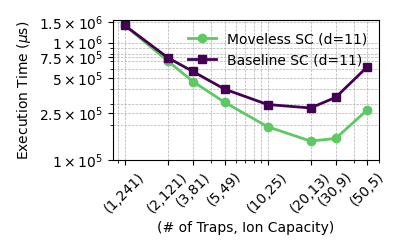}
   \caption{We analyze sensitivity to ion capacities of tightly fitting architectures to the distance 11 surface code (where we vary different amounts of traps to fit this code and have different ion capacities to reach this).}
    \label{fig:ionSensitivity}
\end{figure}

In previous experiments, we had set the ion trap capacity to 5 as a reasonable chain length for near term QCCD hardware. We also gave all compilers the flexibility to map onto $m$ traps which in reality could lead to very sparse traps or many unused traps. In reality, this ion capacity value is flexible across different hardware, and we imagine tighter configurations where traps are forced to be dense are possible. In Figure \ref{fig:ionSensitivity}, we map the distance 11 surface code to the tightest configurations - where the number of traps multiplied by the ion capacity is roughly equivalent to the total amount of ions needed to make the code. We use this to analyze our Moveless compiler's sensitivity to different trap capacities and trap counts.   

\section{Conclusion}

We find that syndrome extraction circuits have large amounts of additional latency due to shuttling operations. This latency translates to an increase in logical error due to amplitude damping and dephasing, and an increased logical error rate. Because of this, we create a compiler we designate as ``Moveless", with policies to minimize intertrap movement in an effort to reduce shuttling operations and lower logical error rate purely through software means. These policies include putting limitations on ancilla qubits to only move ancilla, dynamic scheduling of stabilizers and gates within stabilizers, and reusing ancilla.

Our compiler achieves a best case speedup of $5.24\times$, worst case speedup of $1.4\times$, and the geometrical mean of speedups is of $2.34\times$ across our entire suite of codes when the baseline is given the optimistic assumption of being able to find its optimal ancillary count. This results in lower injected physical error, and demonstrated lower logical error rate (up to 2 orders of magnitude) on simulations of surface and color codes. Our compiler also demonstrates equal or better performance than the baseline with a grid architecture as opposed to linear architectures, and this trend holds irrespective of varying trap counts/ion capacities. Moveless demonstrates how QEC tailored QCCD compilers can achieve lower execution times and lower logical error rates purely through software improvements. Our code is open source on GitHub \cite{movelessgithub}.

\section*{Acknowledgments} This work was funded by the NSF Quantum
Leap Challenge Institute for Robust Quantum Simulation (OMA-2120757) and the NSF STAQ project (PHY-2325080).

%%%%%%%%% -- BIB STYLE AND FILE -- %%%%%%%%
% \clearpage
\bibliographystyle{IEEEtran}
\bibliography{references}
%%%%%%%%%%%%%%%%%%%%%%%%%%%%%%%%%%%%

\end{document}